\documentstyle[sprocl]{article}
\input{psfig}
\newcommand{\onlinecite}{\ref}
\begin{document}

\title{
THE EQUILIBRIUM SHAPE OF InAs QUANTUM DOTS GROWN ON A GaAs\,(001) SUBSTRATE
}

\author{
E. PEHLKE, 
N. MOLL,\footnote{Present address: Massachusetts Institute of Technology,
Department of Physics, Cambridge, MA 02139, USA.}
AND M. SCHEFFLER}

\address{
Fritz-Haber-Institut, 
Faradayweg 4-6, D-14195 Berlin-Dahlem, Germany
}

\maketitle\abstracts{
The equilibrium shape of strained InAs quantum dots grown epitaxially on a 
GaAs(001) substrate is derived as a function of volume. 
InAs surface energies are 
calculated within density-functional theory, and a continuum approach is 
applied for the elastic relaxation energies.
}

\section{Introduction}

The strain-induced self-assembly of small three-dimensional islands 
during heteroepitaxial growth represents an efficient way of producing 
quantum dots.
A frequently studied example is InAs/GaAs(100),\cite{Leonard:93} with the lattice
mismatch amounting to about 7\%. 
InAs grows on GaAs in the Stranski-Krastanov mode: Due to the
smaller surface energy of InAs as compared to GaAs first a
wetting layer forms, but when more InAs is deposited coherent,
dislocation-free three-dimensional islands appear.
The driving force for the island-formation originates in the gain of 
elastic relaxation energy which overcompensates the energetical
costs due to the increased surface area.
However, the details of the growth mechanism, 
especially the reason behind the narrow size distribution found for 
InAs/GaAs(100) quantum dots, are still 
controversial.\cite{Priester:95,Shchukin:95}

In this paper we will derive the equilibrium shape of the InAs quantum dots
as a function of size. With restricting ourselves to the energetics 
we do not imply that kinetic effects are unimportant for shaping
the dots. However, we expect the equilibrium 3D-island shapes to be 
observable under appropriate experimental conditions. When the concentration
of dots is low,  
shape equilibration by atomic diffusion on the small islands will be 
faster than material exchange between the islands 
leading to Ostwald ripening.

To calculate the equilibrium 3D-island shapes we 
have computed InAs surface energies from first-principles for a variety 
of surface orientations. A continuum approach
is adequate to calculate the elastic relaxation energy.
Combining these two contributions we get the total energy, 
which is minimized with respect to the island shape. 
The more delicate energetical effects, like island-island interaction,
the strain-dependence of surface energy etc., 
which have to be considered when discussing island sizes,\cite{Shchukin:95}
can be neglected for the purpose of predicting island shapes at fixed volume.

\section{InAs Surface Energies}

Surface energies have been calculated
using density-functional theory, with the
local-density approximation being applied to the exchange-correlation functional. 
The surface is represented by a periodically repeated slab, 
about 10 atomic layers thick, with 
the topmost 4 layers being fully relaxed. 
The In and As atoms are described by {\it ab initio} 
pseudopotentials,
and the electron density is calculated using 
special {\bf k}-point sets with a
density in reciprocal space equivalent to 64 {\bf k}-points
in the whole (100)\,(1$\times$1) surface Brillouin-zone.
The wave-functions are expanded into plane waves with a 
kinetic energy $\le$10 Ry. 
We have employed a generalized version of the computer 
code {\tt fhi93cp}\,\cite{stumpf:94} to calculate the
energy density according to Chetty and Martin.\cite{Chetty:92} 
The surface energy is derived from integrals of the
energy density over appropriately chosen subvolumes of the
supercell.
For a detailed description of similar calculations for
GaAs see Ref.\,\onlinecite{Moll:96}.

\begin{table}[bht]
\caption{
        The equilibrium surface reconstructions 
        of InAs under As-rich conditions and their surface energies.
        \label{surface_energies}
        }
\vspace{0.2cm}
\begin{center}
\begin{tabular}{|c|c|c|}
\hline
orientation & reconstruction                      & surface energy [meV/\AA$^2$] \\
\hline
(110)       & (1$\times$1) relaxed cleavage plane & 41  \\
(100)	    & $\alpha$(2$\times$4)                & 48  \\
(100)       & c(4$\times$4)                       & 47  \\
(111)	    & (2$\times$2) In vacancy             & 42  \\
(\=1\=1\=1) & (2$\times$2) As trimer              & 36  \\
\hline
\end{tabular}
\end{center}
\end{table}

As the InAs surface reconstructions are expected to be similar to
those of GaAs
we have carried out calculations for the 
same surface reconstructions as in Ref.\,\onlinecite{Moll:96}.
Epitaxial growth most often takes place under As-rich conditions,
thus the data in Table~\ref{surface_energies}
refer to surfaces in equilibrium with bulk As (A7 structure).
As opposed to GaAs the As-terminated InAs(110) (1$\times$1) surface does not 
become stable, instead the relaxed cleavage plane is 
energetically preferred.
For InAs(100) the As-terminated c(4$\times$4) reconstruction yields the lowest
surface energy. However, the energy difference with respect to the $\alpha$(2$\times$4)
reconstruction is so small that our calculation is compatible to the
experimental observation of a (2$\times$4) reconstruction.\cite{Yamaguchi:95} 
Our (111) and (\=1\=1\=1) equilibrium reconstructions are 
consistent with recent core-level and valence-band photoemission 
studies.\cite{Olsson:96,Andersson:96} 
The (111)(2$\times$2) As-trimer reconstruction, which is stable for
GaAs under As-rich conditions, does not become stable in case of InAs.
The surfaces in Table~\ref{surface_energies}
are thermodynamically stable against faceting into each other.
In fact, all four facets have been
observed on large, and thus presumably relaxed, InAs islands
grown on a GaAs(001) substrate.\cite{Steimetz:96}

\begin{figure}[bht]
\centerline{\psfig{figure=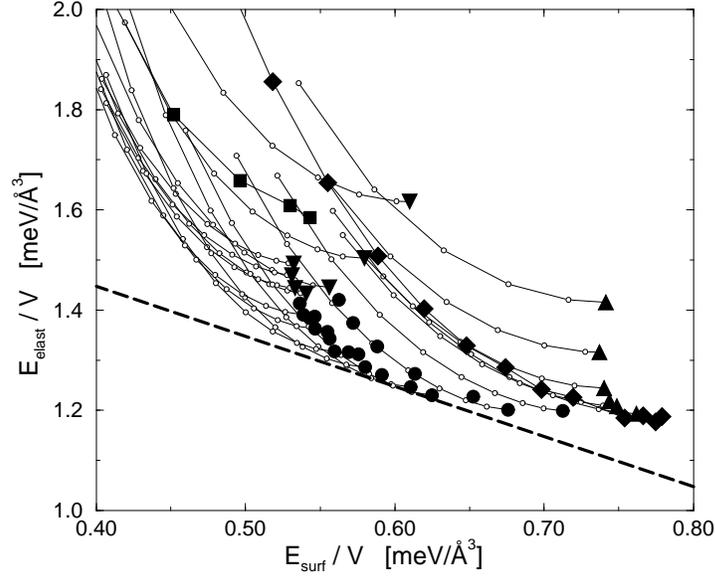,height=7.5cm}}
\caption{Elastic energy per volume $E_{elast}/V$  vs. surface energy per volume
         $E_{surf}/V$ for InAs islands with volume $V=2.88\times 10^5$\AA$^3$.
         Squares: square based pyramid with \{101\} faces and (001)-truncated \{101\}-pyramids.
         Diamonds: square based pyramids with \{111\} and \{\=1\=1\=1\} faces and (001)-truncated pyramids.
         Triangles up: ``huts'' with \{111\} and \{\=1\=1\=1\} faces.
         Triangles down: square based \{101\} pyramids with \{\=1\=1\=1\}-truncated edges.
         Dots: islands with \{101\}, \{111\}, and \{\=1\=1\=1\} faces.
         Filled symbols denote numerical results, while open circles correspond to a
         simple analytical approximation for (001)-truncated ``mesa-shaped'' islands.
         It is assumed that the elastic energy does not change when the
         (almost fully relaxed) top of an island is cut off.
         Full lines connect islands that are created in this way, varying the height of the
         (001) surface plane.
         The dashed line is the curve of constant total energy $E_{elast} + E_{surf}$
         that selects the equilibrium shape.
         \label{Energy}
        }
\end{figure}

\section{Elastic Energy and Equilibrium Shape of Quantum Dots}

The elastic energy is calculated within continuum theory, using a
finite-element type of approach. 
The strain field both in the island and in the substrate is fully 
accounted for.
For simplicity, the same elastic constants\,\cite{cijkl} have been
taken both for the InAs island and the GaAs substrate. 
The results are summarized in Fig.~\ref{Energy} for a specific volume $V$
of the islands. Generalization to arbitrary volume can be done by means of 
the scaling relations $E_{elast}\sim V$ and $E_{surf}\sim V^{2/3}$.
The equilibrium island-shape is marked by the point where the 
lines of constant total energy $E_{elast} + E_{surf}$ 
(dashed line in Fig.~\ref{Energy})
touch the manifold of island energy-curves from below.
The volume-dependence of this shape can be inferred from the same figure 
by rescaling the
$E_{surf}/V$ axis: For larger volume $V$ the slope of the 
total-energy lines becomes smaller.

The equilibrium island shape results from the competition between 
elastic and surface energy. 
According to the different scaling properties of $E_{elast}$ and $E_{surf}$
for large volumes the elastic energy 
dominates, favoring a steep pyramidal shape, while small islands become more and more
flat. Within the configuration space we have examined 
the optimum shapes can be described as mesa-type hills 
bounded by \{101\}, \{111\} and \{\=1\=1\=1\} 
faces and an (001)-surface on the top. While they differ from the \{101\}-pyramids 
seen in experiment~\cite{Ruvimov:95} 
they are in fact similar to shapes
observed for InP/GaInP islands.\cite{Georgsson:95}

>From Fig.~\ref{Energy} it is obvious that the shape 
evolution is continuous with respect to the volume. 
This causes an additional $V$-dependence in
$E_{elast}^{equil}(V)$ and $E_{surf}^{equil}(V)$,
which does not follow ``standard'' scaling anymore. 
Though our total-energy expression does not
yield any optimum island size by its construction, this effect
is important for
a theory of equilibrium island size.\cite{Shchukin:95}

\end{document}